# `TorchSim`: An efficient atomistic simulation engine in PyTorch

**Orion Cohen**[1]   **Janosh Riebesell**[1]   **Rhys Goodall**[1]

**Adeesh Kolluru**[1]   **Stefano Falletta**[1]   **Joseph Krause**[1]

**Jorge Colindres**[1]   **Gerbrand Ceder**[1]   **Abhijeet S Gangan**[1,2,*]

[1]Radical AI, Inc.,
430 E 29th St, Suite 1530,
New York, NY 10016

[2]Department of Civil and Environmental Engineering,
University of California, Los Angeles,
CA 90095, USA

[*]Corresponding author

August 06, 2025

# 1 Abstract

We introduce `TorchSim`, an open-source atomistic simulation engine tailored for the Machine Learned Interatomic Potential (MLIP) era. By rewriting core atomistic simulation primitives in PyTorch [1], `TorchSim` can achieve orders of magnitude acceleration for popular MLIPs. Unlike existing molecular dynamics packages, which simulate one system at a time, `TorchSim` performs batched simulations that efficiently utilize modern GPUs by evolving multiple systems concurrently. `TorchSim` supports molecular dynamics integrators, structural relaxation optimizers, both machine-learned and classical interatomic potentials (such as Lennard-Jones, Morse, soft-sphere), batching with automatic memory management, differentiable simulation, and integration with popular materials informatics tools [2].

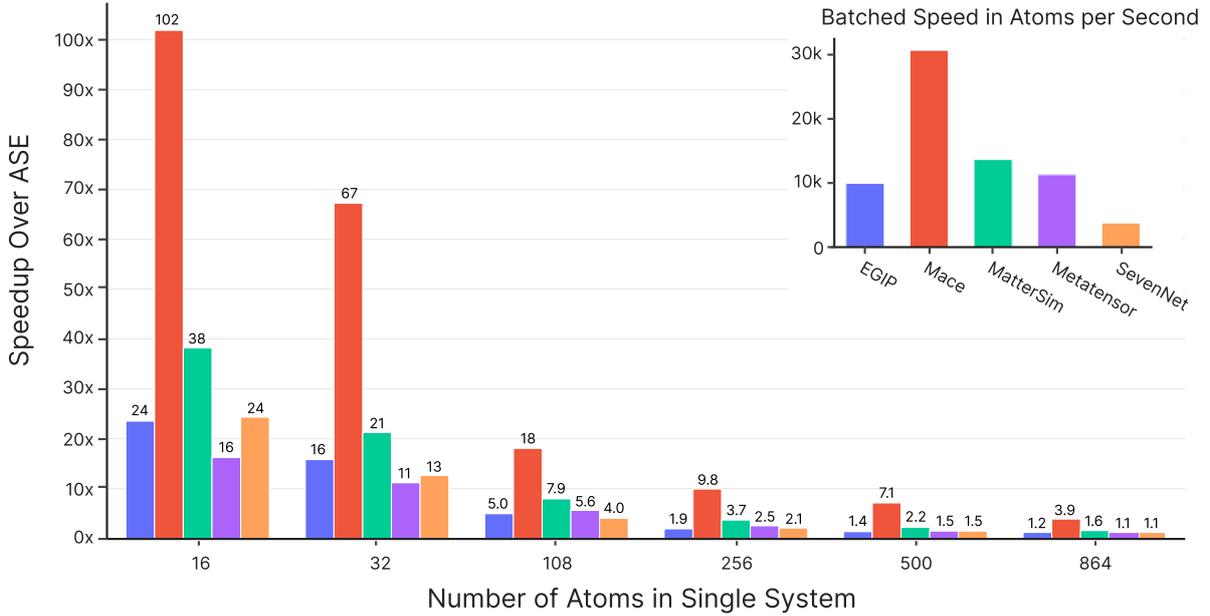

Figure 1: Performance speedup achieved by `TorchSim` for various MLIPs compared to baseline implementations. The inset chart shows the atoms per second for each model with a system size of 864.

## 2 Introduction

In modern materials research, atomistic simulations are indispensable tools complementing theoretical and experimental approaches, forming a crucial pillar of the integrated materials design paradigm. Interatomic potentials commonly simulate atomistic material properties using techniques like molecular dynamics (MD) or structural optimization. Traditional classical interatomic potentials are computationally inexpensive compared to quantum mechanics-based methods, such as Density Functional Theory (DFT) [3], [4]. Although fast, they often lack the transferability and accuracy needed to simulate a wide range of materials and properties. Recent years have seen significant efforts to apply machine learning (ML) methods to construct interatomic potentials. These models, known as Machine-Learned Interatomic Potentials (MLIPs), perform surprisingly well compared to classical potentials, approaching chemical accuracy while offering greater generalizability and transferability [5], [6], [7], [8]. Typically implemented and trained in ML libraries like PyTorch [1], these models heavily utilize GPU hardware. Most traditional atomistic simulation packages (e.g., LAMMPS [9], ASE [10], HOOMD-blue [11]), however, lack native support for GPU operations and/or batched simulations at inference time. As a result, they cannot fully leverage modern hardware.

Table 1: Comparison between `TorchSim` and relevant alternative packages. Green indicates full support, yellow partial support, and red no support. Note this is a snapshot at time of publication. Supported features may change in future versions of these packages.

|  | OpenMM | LAMMPS | TorchMD | ASE | JaxMD | `TorchSim` |
|---|---|---|---|---|---|---|
| **Batching** | ✗ | ✗ | ⊙ | ✗ | ⊙ | ✓ |
| **Diverse MLIPs** | ✗ | ✗ | ✗ | ✓ | ✗ | ✓ |
| **Differentiable** | ✗ | ✗ | ✓ | ✗ | ✓ | ✓ |
| **Pure Python** | ✗ | ✗ | ✓ | ✓ | ✓ | ✓ |
| **GPU Dynamics** | ✓ | ✓ | ✓ | ✗ | ✓ | ✓ |
| **Multi GPU** | ✓ | ✓ | ✗ | ✗ | ✗ | ✗ |

Recently, efforts have emerged to develop MD packages within machine learning frameworks, such as JAX MD [12] and TorchMD [13]. While these packages support GPU-accelerated simulations, they often lack integration with popular materials informatics packages and state-of-the-art MLIP models. Furthermore, their lack of batched simulation capabilities leads to suboptimal performance and GPU utilization on modern hardware; relaxing the largest system on the Materials Project [14] would utilize less than 1% of an H100 GPU.

Designed as a native Python library built on PyTorch, `TorchSim` prioritizes seamless integration with the computational chemistry ecosystem. It directly consumes and converts `pymatgen.core.Structure` [2], `ase.Atoms` [10], and `phonopy.PhonopyAtoms` [15] objects to its internal `SimState` representation, performs arbitrary batched integrations or optimizations on those states, and then re-emits the results to any of the three input formats via `torch_sim.io.state_to_(structure|atoms|phonopy)` conversion functions. This interoperability facilitates `TorchSim`'s use within the larger computational chemistry ecosystem, combining its accelerated simulations with the vast array of available open-source analysis and data management tools.

`TorchSim` supports most popular MLIPs and provides a well-defined interface for easily integrating more. At publication time, `TorchSim` supports MACE [5], MatterSim [6], Fairchem models, Metatensor models [7], and SevenNet [8]. `TorchSim` defines a python interface, `torch_sim.models.ModelInterface`, that enforces the expected input/output schema of energies, forces, and stress with Python's inheritance rules. We took care to make it easy to extend model predictions to additional physical degrees of freedom such as Born charges/polarization, spins/magnetic moments and partial charges which future models may provide [16], [17], [18].

A key advantage of implementing simulation primitives directly in PyTorch is the inherent differentiability offered by its autograd engine, which enables backpropagation on model predictions for arbitrary physical observables computed from an entire simulation. For instance, model parameter gradients can be computed with respect to simulated radial distribution functions at the end of an MD trajectory, then compared to experimental data and used to optimize model parameters towards better agreement with experiment [19]. This opens up exciting possibilities for grounding MLIPs directly in real-world data and potentially training them to better-than-DFT agreement with experiment.

The myriad applications of MLIPs are well covered elsewhere [20]. This work will describe `TorchSim`'s approach to batched simulations, the resulting speedup it attains, and future directions for the package.

# 3 AutoBatching

Batched simulation in `TorchSim` refers to the simultaneous evolution of many disparate systems on a single GPU. The attributes of individual systems, such as positions and momenta, are concatenated into a single `SimState` object that tracks all subsystems. `TorchSim` concatenates atom-wise (positions, velocities, masses, etc.) attributes and stacks batch-wise attributes (unit cell, stress tensor, etc.), enabling it to evolve systems of different sizes together. `TorchSim` batches both the model calls and core mathematical operations of the integrators and optimizers. In `TorchSim`, simulations involving more than one system are always batched. When the cumulative system size exceeds available GPU memory, `TorchSim` automatically organizes the simulations to stay under the memory limit, a feature we call AutoBatching.

Efficiently allocating memory for arbitrary systems and models presents a challenge. Models possess different memory footprints that scale differently, and predicting the maximum system size *a priori* is difficult. For example, the MACE [5] model's memory usage scales with the number of atoms multiplied by the number density (influenced by the radial cutoff), whereas the EGIP [21] model scales only with the number of atoms (due to a maximum neighbor count). Though the memory scaling can be easily inferred from the model architecture, the actual memory usage cannot. Instead, the maximum number of atoms supported must be determined empirically based on the specific hardware.

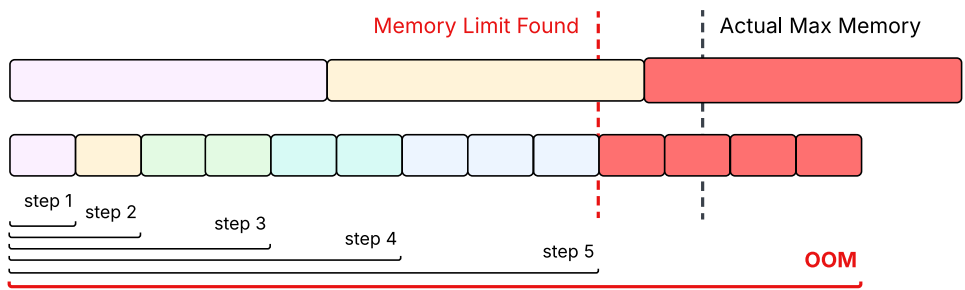

1. Determine memory scalers of systems

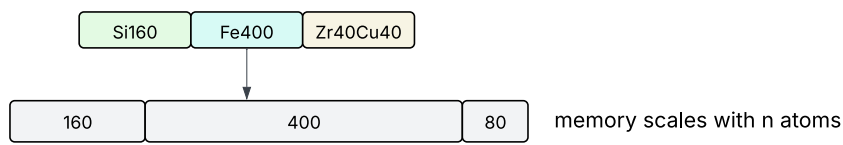

memory scales with n atoms

2. Select largest and smallest systems

3. Iteratively try more systems until OOM

4. Conservatively calculate max scaler

```
max memory scaler 1 = n systems * memory scaler = 2 * 400 = 800

max memory scaler 2 = n systems * memory scaler = 9 * 80  = 720

max memory scaler = min(800, 720) = 720
```

Figure 2: `TorchSim`'s approach to memory estimation illustrated for a batch of three systems using a hypothetical model whose memory footprint scales with the number of atoms (`n_atoms`) and has a maximum memory capacity of approximately 900 atoms. In practice, the actual maximum memory footprint fluctuates depending on the runtime environment.

`TorchSim` provides an automatic memory estimation tool that determines maximum GPU memory by iteratively running larger batches. Each model has a `memory_scales_with` parameter that defines the units for the maximum memory. For models with a fixed max neighbors, this unit is `n_atoms`. For models with a radial neighbor list cutoff, it is `n_atoms_x_density`. `TorchSim` determines the maximum memory by executing forward passes on progressively larger batches until encountering an Out-of-Memory (OOM) error, as depicted in Figure 2. The maximum memory can be determined on-the-fly or calculated once and set manually thereafter.

`TorchSim` supports two approaches to batched simulation, the `BinningAutobatcher` and `InFlightAutobatcher`, which support fixed and variable length simulations, respectively (Figure 3). Fixed-length refers to simulations where the simulation time is identical for many systems and is known in advance, such as running a 100 ps MD simulation on 1000 systems. In this case it is possible to determine the minimum number of total batches by executing a 1D bin-packing algorithm. The memory footprint of each system is calculated in units of `memory_scales_with`, and then the systems are optimally packed into bins of length equal to the maximum memory of the model. The bins, once packed, are evolved one by one until all systems are complete. This ensures optimal use of available memory across a wide range of system sizes. In variable-length simulations, such as geometry relaxation, the simulation time is not known in advance, so a bin packing approach cannot be used. Instead, the `InFlightAutobatcher` monitors the convergence of systems and actively swaps completed systems for new systems. The `InFlightAutobatcher` accepts a stream of incoming systems and, upon initialization, samples systems from the stream until memory is full. After every few steps of optimization, the convergence function is executed on the entire batch. Any converged batches are removed from the optimization and new systems are sampled from the stream until the memory is full. This process repeats until all systems have converged.

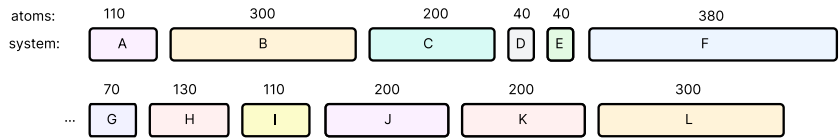
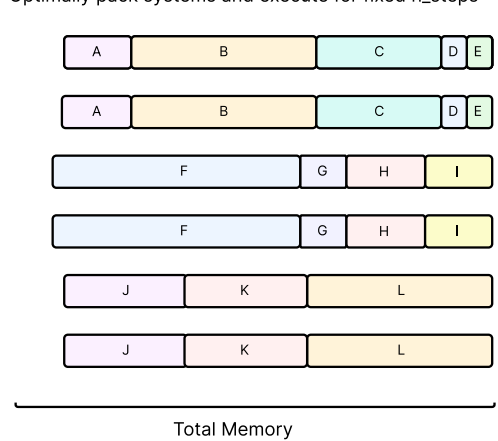
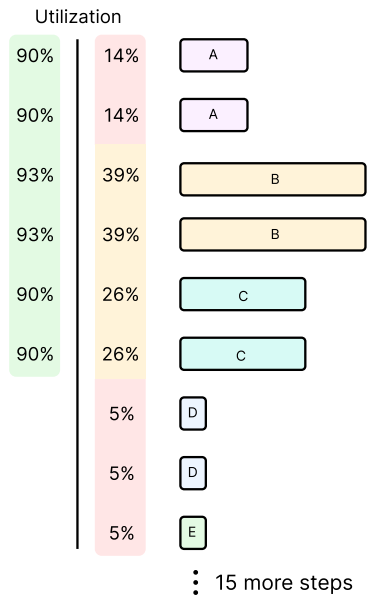
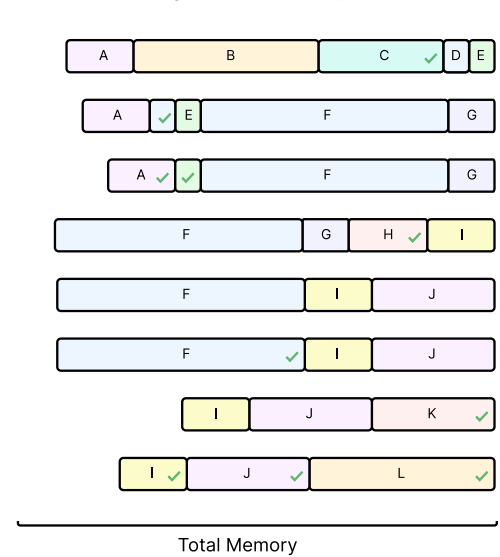
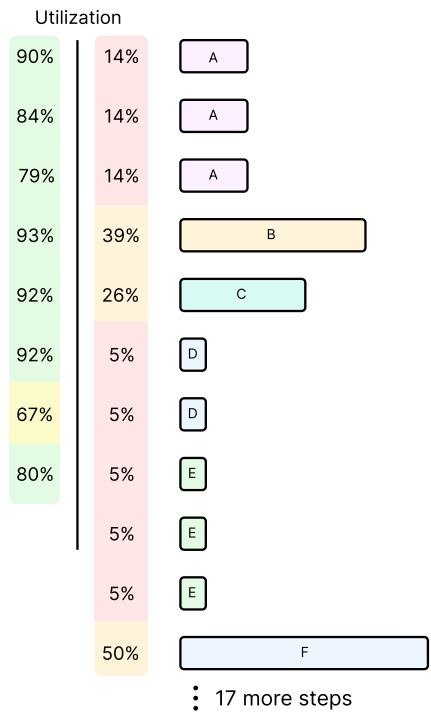

Figure 3: A diagrammatic depiction of the `BinningAutobatcher` and `InFlightAutobatcher` processing a pool of 12 structures. The `BinningAutobatcher` evolves systems for a fixed number of steps, whereas the `InFlightAutobatcher` evolves systems until convergence. The utilization graph shows the memory usage of the batch compared to a hypothetical maximum of 760 atoms.

# 4 Speed Benchmarking

TorchSim achieves an order of magnitude speedup over ASE on small systems for all implemented models (Figure 1). This speedup stems from TorchSim's batched simulation approach, which leverages the available hardware much more efficiently than unbatched simulations. As the system size approaches the memory limit, TorchSim's performance approaches that of serial simulations run in ASE.

We benchmarked `TorchSim`'s performance on face-centered cubic copper systems of 16 to 864 atoms simulated using several methods. Each model's maximum number of atoms was determined by the procedure described in Figure 2. For each system size, we performed simulations with four setups: 1) serial Langevin MD executed in ASE, 2) serial Langevin MD executed in TorchSim, 3) single-step MLIP force evaluations executed in TorchSim, and 4) batched Langevin MD executed in batches in TorchSim. A brief warmup of 15 steps was followed by either 150 steps of MD or 150 forward passes executed in a loop. For the serial systems, batch size was set to one, so the total number of atoms equaled the system size. For batched systems, we set the total number of atoms to `batch_size = max_model_capacity // system_size` – close to the maximum capacity of each model. We ran all benchmarks on an H100, where the maximum number of atoms for the highlighted models was as follows: MACE-MPA-0: 8,000, EGIP: 10,000, SevenNet-MF-OMPA: 3,200, Mattersim V1 1M: 22,000, and PET-MAD: 9,000.

We report model speed in units of atoms per second, which normalizes the influence of system size. Batched simulations interweave the role of memory usage and clock speed. Because more memory-efficient models can fit more batches in memory, they can achieve higher throughput. A model might be fast for small systems but memory-inefficient and unable to scale. This is not reflected in the more traditional metric of nanoseconds per day. Absolute model speeds in atoms/sec are shown in the inset of Figure 1.

In general, the cost of model forward passes was large enough to outweigh both the cost of integration and any differences between serial ASE integration and serial TorchSim integration. Performing looped forward passes was not significantly faster than performing integration with MD, implying that the handful of operations involved in integration were much less expensive than the model calls. We found a similarly insignificant variation in our comparison of serial MD in TorchSim and ASE. Instead, the speedup comes from batching, as shown in Figure 1, which compares serial ASE MD to batched `TorchSim` MD. MACE-MPA-0, for example, could run 500 16-atom systems in parallel with a total throughput 100 times faster than running them serially.

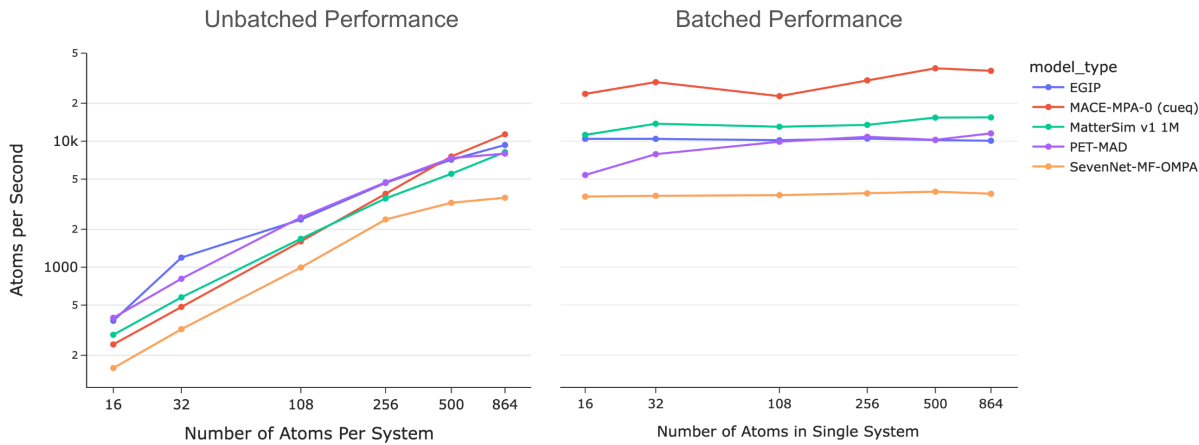

Figure 4: Performance comparison (atoms/second) of various MLIPs running unbatched and batched simulations in `TorchSim` on a single H100 GPU. The x and y axes are log scaled.

Figure 4 shows the atoms per second speed metric for both unbatched and batched integration on a single H100 GPU. The left-hand panel showing the unbatched performance leads to two key conclusions. First, as expected, atoms per second scales slightly sublinearly with system size for unbatched simulations. Second, MACE [5] demonstrates superior scaling compared to other models tested. Although slower for very small systems (e.g., 16 atoms), MACE becomes the fastest model at larger system sizes. In the right-hand panel of Figure 4, we observe that the size of individual systems has a significantly less pronounced effect on inference speed compared to the unbatched serial case. All systems are processed simultaneously in the model forward pass, so the system size becomes mostly irrelevant. We hypothesize that for the different models the inference time remains relatively constant, and the observed speed differences stem primarily from varying graph construction costs.

# 5 Future Work

The future development of `TorchSim` will focus on expanding its capabilities and enhancing the user experience. The roadmap includes deeper integration of batching into common computational materials science workflows like the nudged elastic band method, deeper integration with `phonopy` [15] for increased utilization of batching for running second and third order force constant displacements in parallel, incorporating additional state-of-the-art MLIPs, supporting more complex simulation types (e.g., grand canonical Monte Carlo, metadynamics), improving parallelization strategies for multi-GPU and multi-node execution, and developing enhanced tooling for analysis and visualization. We also aim to further optimize memory management and explore advanced compilation techniques, such as `torch.compile`, to push performance boundaries. We welcome issue reports and community code contributions to the MIT-licensed `TorchSim` GitHub repository at https://github.com/Radical-AI/torch-sim.

# 6 Acknowledgements

This work was supported by Radical AI, Inc. and the University of California, Los Angeles. We thank Kohei Shinohara and Venkat Kapil for their highly valuable discussions regarding algorithm implementation details, API design, and user needs.

# Bibliography

[1] A. Paszke *et al.*, "PyTorch: An Imperative Style, High-Performance Deep Learning Library." Accessed: Aug. 27, 2022. [Online]. Available: http://arxiv.org/abs/1912.01703

[2] S. P. Ong *et al.*, "Python Materials Genomics (Pymatgen): A Robust, Open-Source Python Library for Materials Analysis," *Computational Materials Science*, vol. 68, pp. 314–319, Feb. 2013, doi: 10.1016/j.commatsci.2012.10.028.

[3] G. Kresse and J. Furthmüller, "Efficient Iterative Schemes for Ab Initio Total-Energy Calculations Using a Plane-Wave Basis Set," *Physical Review B*, vol. 54, no. 16, pp. 11169–11186, Oct. 1996, doi: 10.1103/PhysRevB.54.11169.

[4] G. Kresse and J. Furthmüller, "Efficiency of Ab-Initio Total Energy Calculations for Metals and Semiconductors Using a Plane-Wave Basis Set," *Computational materials science*, vol. 6, no. 1, pp. 15–50, 1996.

[5] I. Batatia, D. P. Kovács, G. N. C. Simm, C. Ortner, and G. Csányi, "MACE: Higher Order Equivariant Message Passing Neural Networks for Fast and Accurate Force Fields." Accessed: May 24, 2023. [Online]. Available: http://arxiv.org/abs/2206.07697

[6] H. Yang *et al.*, "MatterSim: A Deep Learning Atomistic Model Across Elements, Temperatures and Pressures." [Online]. Available: https://arxiv.org/abs/2405.04967

[7] A. Mazitov *et al.*, "PET-MAD, a universal interatomic potential for advanced materials modeling." [Online]. Available: https://arxiv.org/abs/2503.14118

[8] Y. Park, J. Kim, S. Hwang, and S. Han, "Scalable Parallel Algorithm for Graph Neural Network Interatomic Potentials in Molecular Dynamics Simulations," *Journal of Chemical Theory and Computation*, vol. 20, no. 11, pp. 4857–4868, May 2024, doi: 10.1021/acs.jctc.4c00190.

[9] A. P. Thompson *et al.*, "LAMMPS - a Flexible Simulation Tool for Particle-Based Materials Modeling at the Atomic, Meso, and Continuum Scales," *Computer Physics Communications*, vol. 271, p. 108171, Feb. 2022, doi: 10.1016/j.cpc.2021.108171.

[10] A. H. Larsen *et al.*, "The Atomic Simulation Environment—a Python Library for Working with Atoms," *Journal of Physics: Condensed Matter*, vol. 29, no. 27, p. 273002, 2017, Accessed: Mar. 01, 2024. [Online]. Available: https://doi.org/10.1088/1361-648X/aa680e

[11] J. A. Anderson, J. Glaser, and S. C. Glotzer, "HOOMD-blue: A Python Package for High-Performance Molecular Dynamics and Hard Particle Monte Carlo Simulations," *Computational Materials Science*, vol. 173, p. 109363, Feb. 2020, doi: 10.1016/j.commatsci.2019.109363.

[12] S. S. Schoenholz and E. D. Cubuk, "JAX, M.D.: A Framework for Differentiable Physics." Accessed: Apr. 23, 2025. [Online]. Available: http://arxiv.org/abs/1912.04232

[13] S. Doerr *et al.*, "TorchMD: A Deep Learning Framework for Molecular Simulations," *Journal of Chemical Theory and Computation*, vol. 17, no. 4, pp. 2355–2363, Apr. 2021, doi: 10.1021/acs.jctc.0c01343.

[14] A. Jain *et al.*, "Commentary: The Materials Project: A Materials Genome Approach to Accelerating Materials Innovation," *APL Materials*, vol. 1, no. 1, p. 11002, Jul. 2013, doi: 10.1063/1.4812323.

[15] A. Togo, "First-Principles Phonon Calculations with Phonopy and Phono3py," *Journal of the Physical Society of Japan*, vol. 92, no. 1, p. 12001, Jan. 2023, doi: 10.7566/JPSJ.92.012001.

[16] S. Falletta *et al.*, "Unified differentiable learning of electric response," *Nature communications*, vol. 16, no. 1, p. 4031, 2025, doi: 10.1038/s41467-025-59304-1.

[17] B. Deng *et al.*, "CHGNet as a Pretrained Universal Neural Network Potential for Charge-Informed Atomistic Modelling," *Nature Machine Intelligence*, vol. 5, no. 9, pp. 1031–1041, Sep. 2023, doi: 10.1038/s42256-023-00716-3.

[18] P. Zhong, D. Kim, D. S. King, and B. Cheng, "Machine Learning Interatomic Potential Can Infer Electrical Response." Accessed: Apr. 08, 2025. [Online]. Available: http://arxiv.org/abs/2504.05169

[19] A. S. Gangan, S. S. Schoenholz, E. D. Cubuk, M. Bauchy, and N. M. A. Krishnan, "Force Field Optimization by End-to-End Differentiable Atomistic Simulation." Accessed: Apr. 24, 2025. [Online]. Available: http://arxiv.org/abs/2409.13844

[20] J. Riebesell *et al.*, "A Foundation Model for Atomistic Materials Chemistry." Accessed: Jan. 02, 2024. [Online]. Available: https://arxiv.org/abs/2401.00096v1

[21] Radical AI, "EGIP: Another step forward in ML-first materials." [Online]. Available: https://www.radical-ai.com/news/EGIP

# Appendix A

## A.1 Dependency Graph (Heat-Colored by Connectedness)

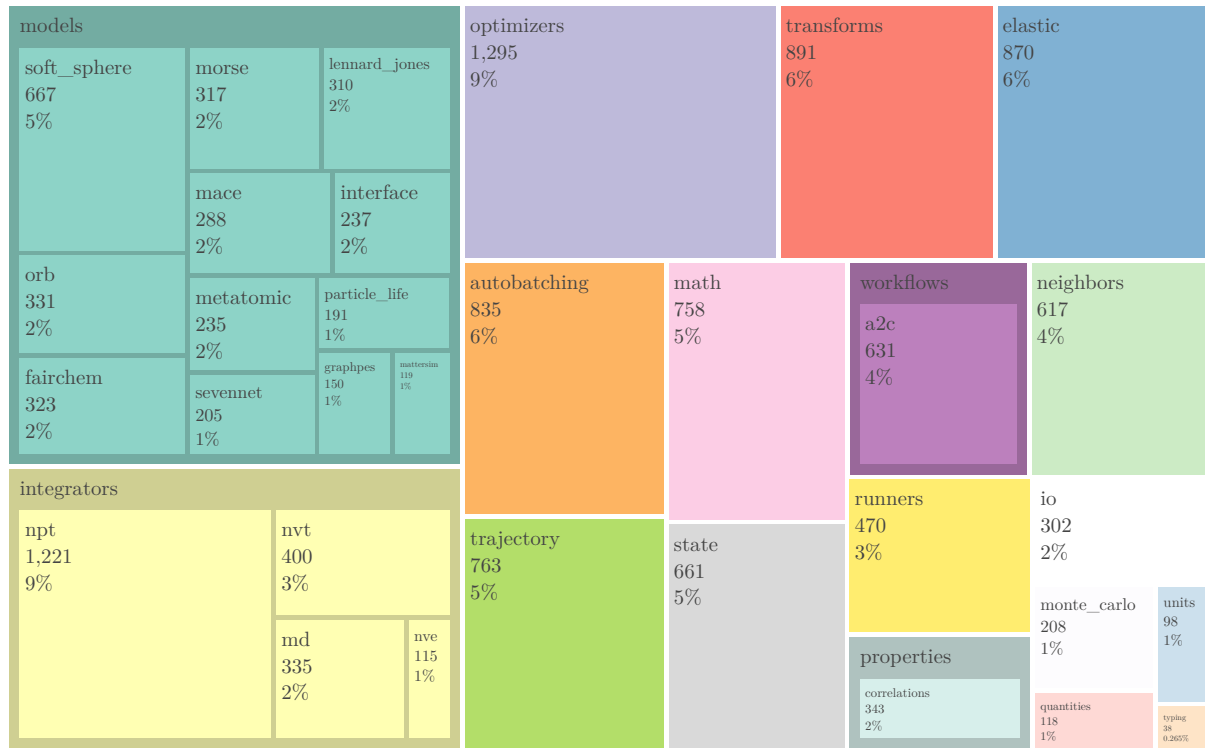

Figure 5: `TorchSim` package structure visualized as a treemap. Each rectangle represents a Python module, displaying its line count and percentage contribution to the total package size.